\documentclass[copyright,creativecommons]{eptcs}
\providecommand{\event}{VPT 2018} %
\usepackage{breakurl}             %
\usepackage{alltt}
\usepackage{amsmath}
\usepackage{amssymb}
\usepackage{color}

\usepackage{tikz} %
\usetikzlibrary{automata} %

\newtheorem{thm}{Theorem}

\newtheorem{dfn}[thm]{Definition} %

\newcommand{\submit}[1]{}

\newcommand{\ignore}[1]{}              
\newcommand{\comment}[1]{}            
\newcommand{\hidecomment}[1]{}         

\newcommand{\ie}{{\it i.e.}}
\newcommand{\eg}{{\it e.g.}}

\newcommand{\wrt}{{with respect to}}

\newcommand{\mymath}[1]{\ensuremath{#1}}

\newcommand{\snode}{\textrm{0}} 
\newcommand{\tnode}{\textrm{1}} 
\newcommand{\Gnterm}{\mathrm{G}}  

\newcommand{\mymid}{\,\,\mid\,\,}
\newcommand{\astfont}{\tt}  

\newcommand{\rawgoto}{\mbox{\astfont goto}}
\newcommand{\rawpop}{\mbox{\astfont pop}}
\newcommand{\rawpush}{\mbox{\astfont push}}
\newcommand{\rawleft}{\mbox{\astfont left}}
\newcommand{\rawright}{\mbox{\astfont right}}
\newcommand{\rawaccept}{\mbox{\astfont accept}}
\newcommand{\rawreject}{\mbox{\astfont reject}}
\newcommand{\rawskip}{\mbox{\astfont skip}}
\newcommand{\rawchoice}{\mbox{\astfont choice}}
\newcommand{\rawchoiceor}{\mbox{\astfont or}}
\newcommand{\rawif}{\mbox{\astfont if}}
\newcommand{\rawthen}{\mbox{\astfont then}}
\newcommand{\rawelse}{\mbox{\astfont else}}
\newcommand{\rawend}{\mbox{\astfont end}}
\newcommand{\rawbottom}{\mbox{\astfont bottom}}
\newcommand{\rawleftend}{\mbox{\astfont leftend}}
\newcommand{\rawrightend}{\mbox{\astfont rightend}}

\newcommand{\rawequal}{\mbox{\astfont =}}
\newcommand{\rawhd}{\mbox{\astfont hd}}
\newcommand{\rawtop}{\mbox{\astfont top}}

\newcommand{\rawcolon}{\mbox{\astfont :}}
\newcommand{\rawsep}{\mbox{\astfont ;}}
\newcommand{\rawnumbase}{\mbox{\astfont 2}} 

\newcommand{\seqast}[2]{{#1}\ {#2}}
\newcommand{\sepast}[1]{{#1}\ \rawsep}
\newcommand{\cmdlabelast}[2]{{#1}\ \rawcolon\ {#2}}
\newcommand{\choiceast}[2]{\rawchoice \ {#1}\ \rawchoiceor\ {#2}\ \rawend}
\newcommand{\ifast}[3]{\rawif \ {#1}\ \rawthen\ {#2}\ \rawelse\ {#3}\ \rawend}
\newcommand{\gotoast}[1]{\rawgoto\ {#1}}

\newcommand{\equalast}[2]{{#1}\ \rawequal\ {#2}}
\newcommand{\popast}{\rawpop}
\newcommand{\pushast}[1]{\rawpush\ {#1}}
\newcommand{\leftast}{\rawleft}
\newcommand{\rightast}{\rawright}
\newcommand{\acceptast}{\rawaccept}
\newcommand{\rejectast}{\rawreject}
\newcommand{\skipast}{\rawskip}
\newcommand{\bottomast}{\rawbottom}
\newcommand{\leftendast}{\rawleftend}
\newcommand{\rightendast}{\rawrightend}

\newcommand{\progv}{Pgm}    
\newcommand{\labelv}{Label} 
\newcommand{\cmdv}{Cmd}     
\newcommand{\cmdsv}{Seq}    
\newcommand{\constv}{Const} 
\newcommand{\bexpv}{Test}   
\newcommand{\sexpv}{Sym}    

\makeatletter
\def\framedstuffsidemargin{1em}
\def\framedstufftopmargin{.7em}
\newenvironment{framedstuff}
  {\setbox42\vbox\bgroup
   \hrule height0pt width\hsize\relax
   \vskip\framedstufftopmargin\relax
   \begin{list}{}
       {\parsep0pt \itemsep1em \parskip0pt
	\leftmargin\framedstuffsidemargin\relax \rightmargin\leftmargin
	\listparindent0pt \itemindent0pt
	\topsep\framedstufftopmargin\relax
	\partopsep\z@\relax
	\labelwidth.3ex}

   \item\relax
  }
  {\par
   \end{list}
   \egroup
   \hrule\hbox to\wd42{\vrule\hskip0ptminus1cm
                       \box42
                       \hskip0ptminus1cm\vrule}\hrule}
\makeatother

\newcommand{\mycaption}[1]{\caption{#1}}

\newcommand{\mypara}[1]{\paragraph{#1.}}

\newcommand{\mypunctspace}{\,}
\newenvironment{codesize}{}{}

\title{
An Experiment in Ping-Pong Protocol Verification \\ 
by Nondeterministic Pushdown Automata}
\author{Robert Gl\"uck
\institute{DIKU, Department of Computer Science, University of Copenhagen}
}
\def\titlerunning{An Experiment in Ping-Pong Protocol Verification 
by Nondeterministic Pushdown Automata}
\def\authorrunning{R.\ Gl\"uck}

\begin{document}
\maketitle

\begin{abstract}

An experiment is described that confirms the security of a well-studied class of cryptographic protocols (Dolev-Yao intruder model) can be verified by two-way nondeterministic pushdown automata (2NPDA). A nondeterministic pushdown program checks whether the intersection of a regular language (the protocol to verify) and a given Dyck language containing all canceling words is empty.
If it is not, an intruder can reveal secret messages sent between trusted users.
The verification is guaranteed to terminate in cubic time at most on a 2NPDA-simulator. The interpretive approach used in this experiment simplifies 
the verification,
by separating the nondeterministic pushdown logic 
and program control, and makes it more predictable. 
We describe the interpretive approach and the known transformational solutions, and show they share interesting 
features. 
Also noteworthy is how abstract results from automata theory can solve practical problems by programming language means. 

\end{abstract}

\section{Introduction}
\label{sec:intro}

Soon after the introduction of public-key encryption~\cite{RSA:78}, it was found that an adversary can obtain a secret message sent on a network between trusted users, not by breaking the cryptographic algorithm, 
but by breaking the communication protocol
through complex interactions with the users. 
A key finding by Dolev and Yao~\cite{DolevYao:81,DolevYao:83} was that the security problem of cryptographic ping-pong protocols can be mapped onto a decidable grammar problem. They gave an algorithm for constructing, for any given ping-pong protocol, a nondeterministic finite-state automaton (regular language) representing all possible interactions between the trusted users and the adversary, and a special-purpose 
algorithm for deciding the security question by computing the collapsing-state relation 
by a closure algorithm.
Their original algorithm decided the security question in time $O(n^8)$, where $n$ is the size of the automaton~\cite{DolevYao:81,DolevYao:83}. This was later improved to $O(n^3)$~\cite{DolevEvenKarp:82}. The security of protocols is very important because public-key crypto systems 
are widely used for electronic communication and underpin various Internet standards.

Recently, Nepeivoda~\cite{Nepeivoda:16} showed that the security of ping-pong protocols can also be verified by program transformation.
Instead of regular expressions, the protocol in question is mapped onto a prefix grammar encoded as a first-order functional program in such a way that it can be used to decide the security question by a program transformer, specifically a 
supercompiler. This 
method 
builds upon 
work solving other verification problems by general-purpose program transformation (\eg,~\cite{AhmedLisitsaNemytykh:13,LisitsaNemytykh:07}). The main steps of these two methods are shown on the left- and right-most branches in Fig.~\ref{fig:VERIFYapproaches}.

This paper takes another programming language approach --- an interpreter for a nondeterministic language is used instead of a program transformer. We show how to write a two-way nondeterministic pushdown (2NPDA) program that searches for insecure communications in a finite-state automaton constructed by the Dolev-Yao algorithm, and interpret the  program by an existing 
simulator for 
nondeterministic pushdown programs, which decides the security question in time $O(n^3)$.
This approach leads to a surprisingly straightforward solution that is asymptotically as efficient as any of the other verification methods above. 
The verification is simplified by separating the problem into an easy-to-write pushdown logic that specifies the solution in terms of 
finding a path 
in a directed graph
and a control component that calculates the actual solution; such separation is known in logic programming as
``algorithm = logic + control''~\cite{Kowalski:79}. 
Herein, 
a nondeterministic pushdown program and a simulator that embeds memoization as a control component are used.
The simulator guarantees termination and polynomial-time performance on a random-access machine with a uniform cost model~\cite{AHU:68,Glueck:16:pdasim}. The simulators that we use simulate one machine by another; in programming language terms these are interpreters.
This experiment
also shows how, relying on proven 
results from automata theory, a 
class of cryptographic protocols
can be verified by means of nondeterministic programming.
The method is shown by the bold shapes in Fig.~\ref{fig:VERIFYapproaches}. The deterministic (det) and nondeterministic (ndet) source and implementation languages of the simulator (SIM) 
and the supercompiler (SCP) are discussed in a later section. The interpretation and transformation approaches are two sides of the same coin, and we show they share interesting features.

\begin{figure}[t]
\begin{framedstuff} %
\vspace{0.05cm}
\newcommand{\mynewline}{\\[-1.8pt]}%
\centering
\begin{minipage}{100mm}
\begin{tikzpicture}[scale=1,>=latex] %

\draw [rounded corners,black,fill=gray!20] (4,7) rectangle 
node[align=center]{ping-pong \mynewline protocol} (6,8);
\draw [->,shorten >= 0.4pt] (5,7) --node[left]{\emph{mapping}~~~~} (3,6);
\draw [->,shorten >= 0.4pt] (5,7) --node[right]{~~~\emph{mapping}} (7,6);

\draw [rounded corners,black,fill=gray!20] (2,5) rectangle 
node[align=center]{finite-state \mynewline automaton} (4,6);
\draw [->,shorten >= 0.4pt] (3,5) --node[left]{\emph{edge set~~~~}} (1,4); %
\draw [->,shorten >= 1.2pt] (3,5) --node[right]{~~~\emph{edges on tape}} (5,4); %

\draw [rounded corners,black,fill=gray!20] (6,5) rectangle
node[align=center]{prefix \mynewline grammar} (8,6);
\draw [->,shorten >= 0.4pt] (7,5) --node[right,align=center]
{~~~~~\emph{program text}}
(9,4);

\draw [rounded corners,black,fill=gray!20] (-0.05,3) rectangle
node[align=center]{state-relation \mynewline closure algo.}
(2.05,4);
\draw [->] (1,3) --node[right]{\emph{answer}}(1,2.05);

\draw [rounded corners,black,fill=gray!20,very thick] (4,3) rectangle 
node[align=center]{pathfinder} 
node[below right]{~~\,\footnotesize ndet} 
(6,4);
\draw [rounded corners,black,fill=gray!20,very thick] (4,2) rectangle
node[]{SIM} 
node[above right]{~~\,\footnotesize ndet}
node[below right]{~~~~\footnotesize det} 
(6,3);
\draw [->,very thick] (5,2) --node[right]{\emph{answer}}(5,1.07);

\draw (5.7,2.8) node[below right]{~~~~\small interpreter}; 
\draw (9.67,3.75) node[below right]{~~~~\small transformer}; 


\draw [rounded corners,black,fill=gray!20] (8,3) rectangle
node[]{SCP} 
node[above right]{~~~~\footnotesize det}
node[below right]{~~~~\footnotesize det} 
(10,4);
\draw [->] (9,3) --node[right]{\emph{answer}}(9,2);

\draw [fill=gray!20] (1,1.5) circle [radius=0.55] 
node[align=center]{y/n};

\draw [fill=gray!20,very thick] (5,0.5) circle [radius=0.57]
node[align=center]{\rule{0cm}{9.5pt}accept \\[-0.6ex] reject}; %

\draw [fill=gray!20] (8.5,1) rectangle
node[align=center]{resid. \mynewline pgm} (9.5,2);
\end{tikzpicture}
\end{minipage}
\end{framedstuff}
\mycaption{Three verification approaches for cryptographic ping-pong protocols: Dolev-Yao's original method~\cite{DolevYao:83} (left) and two programming language approaches --- an interpretive method by pushdown simulation (SIM) described in this paper (in bold) and a transformative method by supercompilation (SCP)~\cite{Nepeivoda:16} (right).
\vspace{-1.00mm}%
}
\label{fig:VERIFYapproaches}
\end{figure}

The approach taken here is similar to other resource-bounded computation models~\cite{Jones:97:complexity} where certain properties are guaranteed for all programs regardless of how they are written (\eg, all programs written in a reversible language are easily invertible~\cite{AxGl11FoSSaCS}). 
Even though the nondeterministic pushdown computation model used here is subuniversal (not Turing-complete), it is not particularly weak. The multihead 2NPDA characterize the polynomial-time (``tractable'') algorithms. Pushdown programming is a technique for solving problems that may deserve more attention, perhaps supported by program transformation.

In Sect.~\ref{sec:pingpong}, we briefly review ping-pong protocols and the security problem. In Sect.~\ref{sec:ndetpgmoverview}, we introduce the nondeterministic pushdown language and, in Sect.~\ref{sec:ndetpgm}, we present the protocol verifier. In Sect.~\ref{sec:vptsim} and~\ref{sec:related}, we discuss the methods and related work, respectively.

\section{Review of Ping-Pong Protocols}
\label{sec:pingpong}

Ping-pong protocols are a class of two-party cryptographic protocols. Their purpose is to transmit secret text between two users in a network. The initiator of a communication applies an initial sequence of operators 
to the text of a
message and sends the message to the intended recipient. 
In each step of their communication, a participant applies an operator sequence to the text most recently received and returns the result. This ping-pong action continues several times as specified by the protocol. Operators that can be applied to a text include name stamps and cryptographic operators~(see~\cite{DolevEvenKarp:82} for more information).

Public-key encryption is a cryptographic system that allows an \emph{encryption key} to be revealed to the public without revealing the corresponding \emph{decryption key} (\eg, RSA~\cite{RSA:78}).
Consequently, a text can be enciphered by anyone using the encryption key publicly revealed by the intended recipient of the text, but only the intended recipient can decipher the text because only this recipient has the corresponding decryption key. There is no need to secretly exchange keys between the participants.
\mypara{Protocol Operators}

A network is assumed to have three legitimate users ($X, Y, Z$) with equal rights. Each user can initiate a communication with another user by sending an initial message.
A message sent in the network consists of three fields: the sender's name, the receiver's name and the text. The text is the part of a message to which a user can apply operators as specified by the communication protocol. 
All users have the same set of operators ($\Sigma$) that they can apply to the text of a message
and each user also has a private operator ($D_X, D_Y, D_Z$) for decrypting a text with their private key.

\begin{dfn}
\label{def:opsets}
The operator sets 
of users $X, Y, Z$ are
\begin{eqnarray}
\Sigma_X &=& \Sigma~\cup~\{ D_X \}~~~~~~~\textrm{decrypt by private key of \mymath{X}}\mypunctspace,\\
\Sigma_Y &=& \Sigma~\cup~\{ D_Y \}~~~~~~~\textrm{decrypt by private key of \mymath{Y}}\mypunctspace,\\
\Sigma_Z &=& \Sigma~\cup~\{ D_Z \}~~~~~~~\textrm{decrypt by private key of \mymath{Z}}\mypunctspace,
\end{eqnarray}
where $\Sigma$ is the common operator set available to every user:
\begin{eqnarray}
\Sigma 
&=& \{E_X, E_Y, E_Z,~~~~~~~~\textrm{encrypt by public key of \mymath{X}, \mymath{Y}, \mymath{Z}} \\
&& ~\;P_X, P_Y, P_Z,~~~~~~~~~~\textrm{prepend name of \mymath{X}, \mymath{Y}, \mymath{Z}} \nonumber\\
&& ~\;M_X, M_Y, M_Z,~~~~~~\textrm{match and delete prepended name of \mymath{X}, \mymath{Y}, \mymath{Z}} \nonumber\\
&& ~\;M \}~~~~~~~~~~~~~~~~~~~~~\textrm{delete any prepended name}\mypunctspace.\nonumber
\end{eqnarray}
\end{dfn}

Operators $E_X, E_Y, E_Z$ encrypt a text with the public key of a user, operators $P_X, P_Y, P_Z$ prepend a user name to a text, operators $M_X, M_Y, M_Z$ delete a prepended user name if the name matches, and $M$ deletes any prepended user name from a text. The cryptographic operators are
defined for any text. 
If the keys mismatch, they return just gibberish (\eg~decryption of an encoded text with the wrong key: $D_Y E_X$).
The cryptographic operators of public-key encryption are inverse to each other (\eg~$D_X E_X = E_X D_X = \epsilon$ where $\epsilon$ denotes the empty sequence of operators). 
A protocol 
aborts 
when prepended names mismatch (\eg~expecting another name stamp: $M_Y P_X$). The order of applying the operators is from right to left. 
\begin{dfn}
\label{def:opids}
Cryptographic operators of public-key encryption~\cite{RSA:78} have the following identities for any user $U$, 
as have the operators for matching and deleting prepended user names~\cite{DolevEvenKarp:82}.
\begin{eqnarray}
\label{eq:identities}
& D_U E_U ~~=~~ E_U D_U ~~=~~M_U P_U  ~~=~~ M P_U ~~=~~ \epsilon\mypunctspace. & %
\end{eqnarray}
\end{dfn}

Consider a Dolev-Yao ping-pong protocol~\cite{DolevEvenKarp:82} as an example that is defined between two users $A,B \in \{ X,Y,Z \}$ where $A$ is the initiator and $B$ the recipient. 
The protocol consists of two operator words, $\alpha_1 \in \Sigma_A^*$ and $\alpha_2 \in \Sigma_B^*$. 
In each step one of the two participants applies an operator word 
to the text and sends the result to the other participant. The following two words define the protocol's ping-pong action.
\begin{eqnarray*}
  \mbox{Protocol 1:}\\
  \alpha_1(A,B) &=& E_B \\
  \alpha_2(A,B) &=& E_A D_B
\end{eqnarray*}
Initially, $A$ sends a message to $B$ after encrypting the secret text by $B$'s public key, that is $A$ applies $\alpha_1(A,B) = E_B$ to the text. Recipient $B$ sends a message back to $A$ after decrypting the received text by $B$'s private key $D_B$ and encrypting the result by $A$'s public key $E_A$, that is $B$ applies $\alpha_2(A,B) = E_A D_B$ to the text it received. The second step completes this ping-pong protocol. The complete operator word applied to the text is $\alpha_2(A,B) \circ \alpha_1(A,B) = E_A D_B E_B$. Under this cryptographic protocol, recipient $B$ can read $A$'s secret text and initiator $A$ can compare the original text with the text echoed by $B$. The communication between $A$ and $B$ is performed without sending any plain text in the network. This cryptographic protocol 
appears to be a secure way of exchanging messages but, as we shall see, it is not.

\begin{figure}[t]
\begin{framedstuff} %
\centering
\begin{minipage}{37mm}
\begin{tikzpicture}[->,>=stealth,       %
                    auto,inner sep=2pt, %
                    semithick,        %
                    node distance=1.1cm,  %
                    xscale=1.1,yscale=1.1] %
\tikzstyle{every node}=[font=\small]
\tikzset{every state/.style={minimum size=5pt}}

\node[state]           (9)  []                 {0};
\node[state,accepting] (1)  [right of=9]       {1};
\node[state]           (2)  [below left  of=1] {};
\node[state]           (3)  [below right of=1] {};

\path (9) edge              node {$E_Y$} (1)
      (1) edge              node {\raisebox{-0.1ex}[0ex][0ex]{$D_Y$}} (3) %
          edge              node {$D_X$} (2)
          edge [loop above] node {$\Sigma_Z$} (1); %

\draw [->,rounded corners] (2) -- node[left]{$E_Y$} (-0.02,-1.57)
                                                    -- (2.6,-1.57) 
                                                    -- (2.6,0)
                                                    -- (1);
\draw [-,rounded corners] (2) -- node[right]{$E_Z$} (0.6,-1.57) -- (0.9,-1.57);
\draw [-,rounded corners] (3) -- node[left] {$E_X$} (1.38,-1.57) -- (1.7,-1.57);
\draw [-,rounded corners] (3) -- node[right]{$E_Z$} (2.0,-1.57) -- (2.2,-1.57);
\end{tikzpicture}
\vspace{0.4ex}%
\end{minipage}
\end{framedstuff}
\mycaption{FSA of Protocol 1.}
\label{fig:FSAprotocolone}
\end{figure}

\mypara{Dolev-Yao Intruder Model}
Assume that two of the users on the network are well-behaved ($X, Y$), which means they only apply the operations specified by the protocol. The third user is a saboteur ($Z$) who can apply any operator sequence $\Sigma_Z^*$ to a message text. 
The saboteur waits patiently for a chance to crack the communication between the well-behaved users by listening to the network, intercepting and altering any message. 
This scenario is sufficient for checking the security of two-party ping-pong protocols in the Dolev-Yao intruder model. 
A single saboteur is sufficient in this model because a single saboteur can do whatever a group of saboteurs can do~\cite{DolevEvenKarp:82}. 
The model assumes only a few limitations on the behavior of the saboteur.
The saboteur can impersonate any sender ($X, Y, Z$), apply any operator sequence $\Sigma_Z^*$ to a text, and send an altered message to any user.
It is assumed that private keys cannot be stolen from the users.

Consider as an example how Protocol 1 can be cracked by saboteur $Z$. Let $X$ 
initiate the communication with $Y$ by sending a text encrypted by $\alpha_1(X,Y)$ in the network. Assume that $Z$ intercepts $X$'s initial message, changes the sender to $Z$, and sends the unchanged text to $Y$, who believes $Z$ initiated a communication by sending the initial $\alpha_1(Z,Y)$. As specified by the protocol, the well-behaved user $Y$ responds to $Z$ by applying $\alpha_2(Z,Y)$ to the received text, that is after decrypting the text by $E_X$ and encrypting it by $E_Z$ for the perceived sender $Z$. The saboteur $Z$ can now simply decrypt the message by~$D_Z$. The secret has been revealed, not by cracking the public-key encryption algorithm, but the protocol! The entire operator sequence applied to the original text is reduced to $\epsilon$ by the operator identities: 
\begin{eqnarray}
& 
D_Z\circ \alpha_2(Z,Y) \circ \alpha_1(X,Y) ~~=~~ 
\fbox{$D_Z E_Z$} \fbox{$D_X E_X$} ~~=~~ 
\epsilon \mypunctspace. 
&
\end{eqnarray}

\mypara{The Security Question}

The Dolev-Yao intruder model formalizes the interaction of the well-behaved users ($X,Y$) and the saboteur ($Z$) using a nondeterministic finite-state automaton (FSA). The FSA for Protocol 1 in Fig.~\ref{fig:FSAprotocolone} generates all operator sequences that can be applied to a text sent by initiator $X$. Assuming that the initiator is $X$ is sufficient for checking the security of the protocol.
The initial state is $\snode$, the accepting state is $\tnode$, and the edges are labeled with operators. 
$X$ starts by sending a message to $Y$ that the saboteur tries to obtain by cracking the protocol.
After the initial operator sequence $\alpha_1(X,Y)= E_Y$ is applied to the text, $Z$ can intercept the message and apply to it an arbitrary operator sequence $\Sigma_Z^*$, or $X$ and $Y$ can apply $\alpha_2(X,Y)$, $\alpha_2(X,Z)$, $\alpha_2(Y,X)$ or $\alpha_2(Y,Z)$ to the text in response to a message received from $X,Y,Z$. The FSA can be constructed for any ping-pong protocol by an algorithm~\cite{DolevEvenKarp:82}.%
 \footnote{We use the reversed and simplified version of the FSA constructed by the algorithm.
}

The security question translates into the following grammar problem: A protocol is secure 
if the intersection of $\mathcal{L}(\mathrm{FSA})$, the regular language defined by its FSA, and $\mathcal{L}(\Gnterm)$, the context-free language of all reducible operator words, is empty. 
Thus, the security question of ping-pong protocols is a decidable grammar problem, namely the emptiness of the intersection of a regular language and a context-language: 
\begin{eqnarray}
\mathcal{L}(\mathrm{FSA}) \cap \mathcal{L}(\Gnterm) &\stackrel{?}{=}& \emptyset\mypunctspace.
\end{eqnarray}
The context-free grammar $\Gnterm$ generating all reducible words is the same for all ping-pong protocols:
\begin{eqnarray}
\Gnterm &::=& D_U~\Gnterm~E_U~\mid~E_U~\Gnterm~D_U~\mid~M_U~\Gnterm~P_U~\mid~M~\Gnterm~P_U~\mid~\Gnterm~\Gnterm~\mid~\epsilon~~~~~\mbox{for all $U \in \{X,Y,Z\}$\mypunctspace.}
\end{eqnarray}
The saboteur 
in the Dolev-Yao intruder model 
can apply any operator sequence $\Sigma_Z^*$ to a message text, so $\Gnterm$ must generate all reducible words.
A word in $\Gnterm$ is balanced with respect to ``opening'' and ``closing'' pairs of operators. 
Any word in $\mathcal{L}(\Gnterm)$ can be \emph{reduced} to $\epsilon$ by repeatedly applying the identity rules in Def.~\ref{def:opids}, {\ie} by substituting $\epsilon$ successively for every occurrence of pairs to which the identity rules apply. 
It can easily be shown that 
these reductions can be performed in any order without changing the result. The \emph{reduction strategy} we are going to use later is to repeatedly apply the identity rules to the rightmost, innermost pair to which they apply.%
  \footnote{$\mathcal{L}(\Gnterm)$ is an ambiguous 
  language: parenthesizing is not necessarily unique,
  {\eg}~\texttt{\fbox{$E_X$\fbox{$D_X E_X$}$D_X$}} or 
  \texttt{\fbox{$E_X D_X$}\fbox{$E_X D_X$}}\mypunctspace.}
There is no difference between handling a mismatch of keys and of
prepended names (the operator pair does not reduce). 

\begin{dfn}
A ping-pong protocol is \emph{secure} iff there is no 
accepting path 
in 
its FSA representation whose word is reducible to $\epsilon$ by the operator identities; otherwise, the protocol is insecure~\cite{DolevEvenKarp:82}. 
\end{dfn}

\begin{figure}[t]
\begin{framedstuff} %
\centering
\begin{minipage}{128.5mm}
\begin{tikzpicture}[baseline=(current bounding box.north),
                    ->,>=stealth,       %
                    auto,inner sep=2pt, %
                    semithick,        %
                    node distance=1.05cm,  %
                    xscale=1.05,yscale=1.05] %
\tikzstyle{every node}=[font=\small]
\tikzset{every state/.style={minimum size=5pt}}       
\node[state]           (8)                    {0};
\node[state]           (9)  [right of=8]       {};
\node[state,accepting] (1)  [right of=9]       {1};
\node[state]           (3)  [below right of=1] {};
\node[state]           (6)  [below of=3]       {};
\node[state]           (7)  [right       of=6] {};
\node[state]           (2)  [below left  of=1] {};
\node[state]           (5)  [below       of=2] {};
\node[state]           (4)  [left        of=5] {};
\path %
      (8) edge node {$P_X$} (9)
      (9) edge node {$E_Y$} (1)
      (1) edge [] node {\raisebox{-0.25ex}[0ex][0ex]{\hspace{0.3ex}$D_Y$}} (3)
          edge [] node {$D_X$} (2)
          edge [loop above] node {$\Sigma_Z$} (1) %
      (2) edge [above left] node {$M_Y$} (4)
          edge [right]      node {$M_Z$} (5)
      (3) edge [left]       node {$M_X$} (6)
          edge              node {$M_Z$} (7);
\draw [->] (4) [rounded corners] -- node[left]{$E_Y$} (0.3,-2.57)
                                                    -- (4.5,-2.57) 
                                                    -- (4.5,0)
                                                    -- (1);
\draw [-] (5) [rounded corners] -- node[right]{$E_Z$} (1.3,-2.57) -- (1.5,-2.57);
\draw [-] (6) [rounded corners] -- node[left] {$E_X$} (2.7,-2.57) -- (2.9,-2.57);
\draw [-] (7) [rounded corners] -- node[right]{$E_Z$} (3.7,-2.57) -- (3.9,-2.57);
\end{tikzpicture}
\hspace{9.0ex}%
\begin{tikzpicture}[baseline=(current bounding box.north),
                    ->,>=stealth,       %
                    auto,inner sep=2pt, %
                    semithick,        %
                    node distance=1.05cm,  %
                    xscale=1.05,yscale=1.05] %
\tikzstyle{every node}=[font=\small]
\tikzset{every state/.style={minimum size=5pt}}    
\node[state]           (0)                     {0}; %
\node[state]           (8)  [right of=0]       {};
\node[state]           (9)  [right of=8]       {};
\node[state,accepting] (1)  [right of=9]       {1};
\node[state]           (3)  [below right of=1] {};
\node[state]           (6)  [below       of=3] {};
\node[state]           (10) [below       of=6] {};
\node[state]           (7)  [right       of=6] {};
\node[state]           (13) [below       of=7] {};
\node[state]           (2)  [below left  of=1] {};
\node[state]           (5)  [below       of=2] {};
\node[state]           (12) [below       of=5] {};
\node[state]           (4)  [left        of=5] {};
\node[state]           (11) [below       of=4] {};
\path (0) edge node {$E_Y$} (8)
      (8) edge node {$P_X$} (9)
      (9) edge node {$E_Y$} (1)
      (1) edge [] node {\raisebox{-0.25ex}[0ex][0ex]{\hspace{0.3ex}$D_Y$}} (3)
          edge [] node {$D_X$} (2)
          edge [loop above] node {$\Sigma_Z$} (1) %
      (3) edge [left] node {$M_X$} (6)
          edge node {$M_Z$} (7)
      (6) edge [left] node {$D_Y$} (10)
      (7) edge node {$D_Y$} (13)
      (2) edge [above left] node {$M_Y$} (4)
          edge [right] node {$M_Z$} (5)
      (5) edge node {$D_X$} (12)
      (4) edge [left] node {$D_X$} (11); 
\draw [->] (11) [rounded corners] -- node[left]{$E_Y$} (1.3,-3.57)
                                                    -- (5.5,-3.57) 
                                                    -- (5.5,0)
                                                    -- (1);
\draw [-] (12) [rounded corners] -- node[right]{$E_Z$} (2.3,-3.57) -- (2.5,-3.57);
\draw [-] (10) [rounded corners] -- node[left]{$E_X$} (3.7,-3.57) -- (3.9,-3.57);
\draw [-] (13) [rounded corners] -- node[right]{$E_Z$} (4.7,-3.57) -- (4.9,-3.57);
\end{tikzpicture}
\vspace{0.4ex}%
\end{minipage}
\end{framedstuff}
\mycaption{FSA of Protocol 2 (secure) and Protocol 3 (insecure).}
\label{fig:FSAprotocoltwothree}
\end{figure}

Protocol 1 can be made secure by prepending the name of initiator $A$ to the text before encrypting it with $B$'s public key, that is $A$ applies $\alpha_1(A,B) = E_B P_A$ to the text. Recipient $B$ now encrypts using $A$'s public key only if the text has $A$'s name prepended, as checked by match $M_A$, that is $B$ applies $\alpha_2(A,B) = E_A M_A D_B$ to the text it received. The FSA of Protocol 2 in Fig.~\ref{fig:FSAprotocoltwothree} has no accepting path whose word is reducible~\cite{DolevEvenKarp:82}. 
\begin{center}
\vspace{-3ex}%
\begin{minipage}[b]{65mm}
\begin{eqnarray*}
\mbox{Protocol 2:} \\
  \alpha_1(A,B) &=& E_B P_A \\
  \alpha_2(A,B) &=& E_A M_A D_B
\end{eqnarray*}
\end{minipage}
\hspace{10mm}
\begin{minipage}[b]{65mm}
\begin{eqnarray*}
\mbox{Protocol 3:} \\
  \alpha_1(A,B) &=& E_B P_A E_B \\
  \alpha_2(A,B) &=& E_A D_B M_A D_B
\end{eqnarray*}
\end{minipage}
\end{center}
One may think that extra encryption makes a protocol even more secure, but this is not necessarily the case.
Suppose we want to improve Protocol~2 by encrypting the text once more. Let initiator $A$, before prepending the name to the text by $P_A$, encrypt it with $B$'s public key, that is $\alpha_1(A,B) = E_B P_A E_B$. Receiver $B$ now decrypts the text before encrypting it with $A$'s public key, that is $\alpha_2(A,B) = E_A D_B M_A D_B$. This innocent looking ``improvement'' makes the protocol insecure! This shows that informal arguments are error prone. A formal way to decide the security of ping-pong protocols is needed. 
The operators which a saboteur can inject into the communication are not immediately obvious in Fig.~\ref{fig:FSAprotocoltwothree} in the FSA of Protocol~3~\cite{DolevEvenKarp:82}. 
For example, it can be cracked by a patient saboteur who twice injects operators and fakes senders and receivers:
\begin{eqnarray}
D_Z \circ \alpha_2(Z,Y) \circ E_Y P_Z M_X D_Z \circ \alpha_2(Z,Y) \circ E_Y P_Z \circ \alpha_1(X,Y) &=& \nonumber\\
\fbox{$D_Z E_Z$} \fbox{$D_Y \fbox{$M_Z \fbox{$D_Y E_Y$} P_Z$} \fbox{$M_X \fbox{$D_Z E_Z$} \fbox{$D_Y \fbox{$M_Z \fbox{$D_Y E_Y$} P_Z$} E_Y$} P_X$} E_Y$} &=&
\epsilon\mypunctspace.
\end{eqnarray}

\section{Nondeterministic Programming and a Pushdown Language}
\label{sec:ndetpgmoverview}

\emph{Two-way nondeterministic pushdown automata} (2NPDA) will be our programming model.
Pushdown automata are simple versatile devices comprising three components: a read-only input tape, a potentially infinite stack, and a finite-state control. They can move the head on the tape in both directions, push and pop symbols to and from the stack, and test the symbol on top of the stack and the symbol read on the tape.
The alphabets of tape symbols, stack symbols and states are finite.
\emph{Multihead pushdown automata} can read and move multiple heads independently on the tape. 

Even though these devices are subuniversal, they are not particularly weak. The \emph{multihead 2NPDA} 
are equivalent to the
polynomial-time algorithms~\cite{WagnerWechsung:86}. Any $k$-head 2NPDA can be simulated in at most $O(n^{3k})$ steps on a random-access machine with a uniform cost model where $n$ is the length of the tape~\cite{AHU:68}. We shall see that a single head ($k=1$) is sufficient to check the security of ping-pong protocols, which means this takes at most cubic time.

Instead of using a traditional multi-valued transition function for defining the operation of a 2NPDA, 
we introduce the 
\emph{multihead nondeterministic pushdown language}, shown in Fig.~\ref{fig:pdafcl-syntax}, with commands to move the head (\texttt{left}, \texttt{right}), push and pop symbols (\texttt{push}, \texttt{pop}), and halt in an accepting or rejecting final state (\texttt{accept}, \texttt{reject}). Predicates can compare two symbols (\texttt{=}), test the emptiness of the stack (\texttt{bottom}) and the ends of the tape (\texttt{leftend}, \texttt{rightend}). Global variables contain the symbol currently on top of the stack (\texttt{top}) and read by the tape head (\texttt{hd}). The variables are updated when the stack top changes or the head moves. 
Similarly, additional tape heads (\texttt{hd2}, \texttt{hd3}, $\ldots$) can be moved (\texttt{left2}, \texttt{left3}, $\ldots$) and tested (\texttt{leftend2}, \texttt{leftend3}, $\ldots$). The deterministic control-flow operators are as usual (\texttt{if}, \texttt{goto}). A program consists of labeled command sequences. Execution begins at the first command of a program. 
The left and right ends of the tape are marked with the tape symbols \texttt{>} and \texttt{<}, respectively.

The language resembles flowchart languages except for its additional \texttt{choice} command~\cite{Jones:97:complexity} which nondeterministically executes either command sequence $\cmdsv_1$ or $\cmdsv_2$, that is, the next state after the choice is not uniquely determined by the current state:
\begin{center}
\texttt{choice }$\cmdsv_1$\texttt{ or }$\cmdsv_2$\texttt{ end}
\end{center}

\begin{figure}[t]
\begin{framedstuff}
\begin{center}
\begin{minipage}[t]{80mm}
\tt
\begin{array}[t]{lcl}
\progv &::= & (\cmdlabelast{\labelv}{\cmdsv})^+\\
\cmdsv &::= & \cmdv \mymid 
              \seqast{\sepast{\cmdv}}{\cmdsv}\\
\bexpv &::= & \bottomast \\
       &\mid& \leftendast~\mymid \leftendast\astfont{\rawnumbase}~\mymid \ldots\\
       &\mid& \rightendast\mymid \rightendast\astfont{\rawnumbase}\mymid \ldots\\
       &\mid& \equalast{\sexpv}{\sexpv} \\
\sexpv & ::= & \constv \mymid \rawtop \mymid 
               \rawhd  \mymid \rawhd\astfont{\rawnumbase} \mymid \ldots\\
\end{array}
\end{minipage}
\begin{minipage}[t]{80mm}
\tt
\begin{array}[t]{lcl}
\cmdv  &::= & \popast~~\mymid \pushast{\sexpv}\\
       &\mid& \leftast~\mymid \leftast\astfont{\rawnumbase}~\mymid \ldots\\ 
       &\mid& \rightast\mymid \rightast\astfont{\rawnumbase}\mymid \ldots\\ 
       &\mid& \choiceast{\cmdsv}{\cmdsv}\\
       &\mid& \ifast{\bexpv}{\cmdsv}{\cmdsv}\\
       &\mid& \gotoast{\labelv} \mymid \skipast\\
       &\mid& \acceptast        \mymid \rejectast\\
\end{array}
\end{minipage}
\end{center}
\end{framedstuff}
\mycaption{Syntax of the multihead nondeterministic pushdown language.} %
\label{fig:pdafcl-syntax} %
\end{figure}

Initially, the stack is empty and the tape heads scan the left end of the tape containing the input word. An input word is \emph{accepted} by a nondeterministic pushdown program if there exists
at least one computation sequence for the program that terminates in an \texttt{accept} command. 
The nondeterministic operation allows the simultaneous construction of every computation sequence for a given input.
The formal language accepted by a program is the set of all input words it accepts. This is the usual definition for nondeterministic pushdown automata. The semantics of the pushdown language will not be formally defined here due to lack of space and the semantics being straightforward.

A program that contains no \texttt{left} command is \emph{one way}; one that contains no \texttt{choice} command is \emph{deterministic}. One-head programs accept important classes of formal languages. For example, the one-way nondeterministic pushdown (1NPDA) programs accept the context-free languages.

A textbook interpretation of a pushdown program may 
not terminate 
(\eg, push forever on the stack)
or take exponential time before terminating. However, memoizing simulation methods ensure termination and polynomial-time performance for all pushdown programs, because the number of possible surface configurations is polynomially bounded. Thus, every pushdown program has a definite answer (accept, reject). 
We refer the reader to~\cite{AHU:68,Glueck:16:pdasim} for a presentation of the simulation methods, and to~\cite{Cook:72,Jones:77} for the deterministic case. Accordingly, we shall be programming in a resource-bounded and decidable nondeterministic programming language, following the approach marked bold in Fig.~\ref{fig:VERIFYapproaches}.

\section{Protocol Security Checked by Nondeterministic Pushdown Programs}
\label{sec:ndetpgm}

Combinatorial search problems may often be simply written using nondeterministic programs. Before we show how to verify the security of ping-pong protocols, we show how to find a path between two nodes in a directed graph by a pushdown program. We then extend the pathfinding program into the desired protocol verifier, discuss the nondeterministic programs and report on simulation results for the Dolev-Yao protocols. Verification by program transformation is discussed in the next section.

\subsection{Pathfinding in a Directed Graph by a Pushdown Program} %

\begin{figure}[t]
\begin{framedstuff}
\vspace{0.05cm}
\centering
\begin{minipage}[t]{140mm}
\begin{codesize}
\begin{alltt}
init: push '0'; right            (* push start node, pos 1st edge  *)
loop:	if top = hd                (* both nodes match?              *)
      then choice                (* make a guess: traverse or skip *)
             pop; right;         (* traverse edge                  *)
             if hd = '1' then accept end;  (* path 0 -> 1 found    *)
             push hd; right      (* push next node, pos next edge  *)
           or
             2-right end;        (* skip edge                      *)
      else 2-right end;          (* mismatch: skip edge            *)
      if rightend then move-to-leftend end;  (* return to 1st edge *)
      goto loop
\end{alltt}
\end{codesize}
\end{minipage}
\end{framedstuff}
\mycaption{Finding a path in a directed graph by a nondeterministic PDA-program. The graph is given by its edges on the input tape.}
\label{fig:PDApathfind}
\end{figure}%

Given a \emph{directed graph} $G = (V, E)$ with nodes $u,v \in V$ and directed edges $E = \{(u_1,v_1), \ldots, (u_n,v_n)\}$, where an edge $(u_i,v_i)$ leads from node $u_i$ to node $v_i$, the task of the pushdown program is to check whether there exists a path from a source node $s \in V$ to a target node $t \in V$ in $G$.

Let the names of the source and target nodes 
be fixed as $s=\snode$ and $t=\tnode$. For simplicity, we assume that the node names in $V$ are included in the tape- and stack-symbol alphabets of the  automaton.
$G$ can be represented by a tape of length $O(n)$ that lists all edges in $E$, where \texttt{>} and \texttt{<} mark the two tape ends:
\[
\texttt{>}~u_1~v_1~\ldots~u_n~v_n~\texttt{<}
\]

The nondeterministic program shown in Fig.~\ref{fig:PDApathfind} guesses a path in $G$ from $\snode$ to $\tnode$, if it exists. Initially, the stack is empty and the head is positioned at the left end (\texttt{>}) of the tape, so $\snode$ is pushed on the empty stack by \texttt{push~'0'} and the head is positioned at $u_1$ of the first edge $(u_1~v_1)$ on the tape by \texttt{right} in the first line of the program. The invariant of the main loop that follows, is that the current node of the path that the program is exploring is kept on top of the stack. This node is updated by the main loop when an edge is traversed to a new node. No other nodes are pushed on the stack, so a stack of height 1 suffices for following a path in $G$. Furthermore, a single head suffices for scanning the edges on the tape.

The main loop moves the head \texttt{hd} over the sequence of edges on the tape,  until an edge originating in the current node
on top of the stack 
is found (\texttt{top}~=~\texttt{hd}). The nondeterministic choice at this point is to either traverse the edge originating in \texttt{top} or to continue the search for another edge originating in \texttt{top}.
If the edge $(\texttt{top},v)$ is traversed and $v=\tnode$, then a path from $\snode$ to $\tnode$ exists and the computation halts with \texttt{accept}; otherwise, 
the search continues with $v$ as the new current node. When the right end (\texttt{<}) of the tape is reached during the search, that is predicate \texttt{rightend} is true, the head is repositioned at the first edge by \texttt{move-to-leftend}, and the search continues with the first edge. The main loop cycles over the edges on the tape traversing or skipping edges nondeterministically. An edge takes two positions on the tape, so two right moves skip an edge (shorthand notation \texttt{2-right}). 

Command \texttt{move-to-leftend} is assumedly built-in. It can be implemented by commands
\begin{center}
\begin{minipage}[t]{148mm}
\begin{codesize}
\begin{alltt}
repeat: left; if leftend then right else goto repeat end
\end{alltt}
\end{codesize}
\end{minipage}
\end{center}

The nondeterministic logic of finding a path in a directed graph is straightforward: Follow all paths starting from $\snode$ and accept if $\tnode$ is reached. The nondeterministic program describes how to follow all paths in a pushdown computation model without concern for efficiency and termination (\eg, cycles in graph). The control is separate (in the simulator).
Both of the simulation methods for nondeterministic pushdown automata~\cite{AHU:68,Glueck:16:pdasim} perform a universal search in the space of nondeterministic computations (find all computation sequences), even though an existential search (halt after first accept) decides the problem. Both use a control component consisting of memoization 
for avoiding redundant computations.
Together, the  
pushdown
program and the simulator constitute the algorithm for finding a path 
in a directed~graph.

\subsection{Protocol Security Checking by Pathfinding}

\begin{figure}[t]
\begin{framedstuff}
\vspace{0.05cm}
\centering
\begin{minipage}[t]{148mm}
\begin{codesize}
\begin{alltt}
init: push '0'; right               (* push start node, pos 1st edge   *)
loop:	if top = hd                   (* both nodes match?               *)
      then choice                   (* make a guess: traverse or skip  *)
             pop; right;            (* traverse edge, check identities *)
             if (hd = 'DX' \mymath{\wedge} top = 'EX') \mymath{\vee} (hd = 'EX' \mymath{\wedge} top = 'DX') \mymath{\vee}
                (hd = 'DY' \mymath{\wedge} top = 'EY') \mymath{\vee} (hd = 'EY' \mymath{\wedge} top = 'DY') \mymath{\vee}
                (hd = 'DZ' \mymath{\wedge} top = 'EZ') \mymath{\vee} (hd = 'EZ' \mymath{\wedge} top = 'DZ') \mymath{\vee}
                (hd = 'MX' \mymath{\wedge} top = 'PX') \mymath{\vee} (hd = 'M'  \mymath{\wedge} top = 'PX') \mymath{\vee}
                (hd = 'MY' \mymath{\wedge} top = 'PY') \mymath{\vee} (hd = 'M'  \mymath{\wedge} top = 'PY') \mymath{\vee}
                (hd = 'MZ' \mymath{\wedge} top = 'PZ') \mymath{\vee} (hd = 'M'  \mymath{\wedge} top = 'PZ')
             then pop               (* reduce operator pair to \mymath{\epsilon}       *)
             else push hd end;      (* trace unreducible operator      *)
             right;                 (* move to next node               *)
             if hd = '1' \mymath{\wedge} bottom then accept end;  (* insecure path   \!*)
             push hd; right         (* push next node, pos next edge   *)
           or
             3-right end;           (* skip edge                       *)
 	    else 3-right end;             (* mismatch: skip edge             *)
 	    if rightend then move-to-leftend end;  (* return to 1st edge     *)
 	    goto loop
\end{alltt}
\end{codesize}
\end{minipage}
\end{framedstuff}
\mycaption{Protocol security checking 
by a nondeterministic 2-way PDA-program. The operator-labeled directed graph  representing the security problem of the protocol is given on the input tape.
}
\label{fig:PDAverify}
\end{figure}%

To verify ping-pong protocols, we extend the nondeterministic pathfinding program~(Fig.~\ref{fig:PDApathfind}) to operator-labeled graphs that represent the security problem of a ping-pong protocol (\eg, Figs.~\ref{fig:FSAprotocolone} and~\ref{fig:FSAprotocoltwothree}).

Given an \emph{operator-labeled directed graph} $G = (V, E)$ with nodes $u,v \in V$ and directed edges $E = \{(u_1,o_1,v_1),$ $\dots,$ $(u_n,o_n,v_n)\}$ labeled with the operators 
$o_i \in \Sigma_{\mathit{XYZ}} = \Sigma_X \cup \Sigma_Y \cup \Sigma_Z$ of Def.~\ref{def:opsets}, the task of the pushdown program is to decide whether there exists a path from a source node $s \in V$ to a target node $t \in V$ in  $G$ along which the 
operator word is reducible to $\epsilon$ by the operator identities of Def.~\ref{def:opids}.

As above, let the names of the source and target nodes be fixed as $s=\snode$ and $t=\tnode$. Assume the node names in $V$ and the operator names in $\Sigma_{\mathit{XYZ}}$ are included in the tape- and stack-symbol alphabets of the automaton.
$G$ can be represented by a tape of length $O(n)$ that lists all edges in $E$: 
\[
\texttt{>}~u_1~o_1~v_1~\ldots~u_n~o_n~v_n~\texttt{<}
\]
An edge is represented by three symbols on the tape.
As with pathfinding, the main loop of the pushdown program cycles over the edges on the tape and guesses a path from $\snode$ to $\tnode$, if it exists. In addition, the program collects the operators labeling the edges along the path and tries to reduce the operator word to $\epsilon$ by applying the operator identities.
The program is shown in Fig.~\ref{fig:PDAverify}.
The main parts are as follows:
\begin{itemize}
\item Main loop with nondeterministic choice to guess a path.
\item Applying the operator identities.
\end{itemize}

The current node of the path being explored is kept on top of the stack while searching for matching edges. 
The operators 
that could not be reduced so far along the current path are on the stack below the current node.
The task of the main loop is to find edges originating in the current node by cycling over the edges on the tape.  
An edge takes three positions on the tape, so three right moves skip an edge (shorthand notation 3-right).
When an edge $(\texttt{top},o,v)$ originating in the current node \texttt{top} is nondeterministically selected for traversal by \texttt{choice}, the current node is popped and an attempt is made to reduce the operator~$o$ labeling the selected edge and the last unreduced operator $o_\textit{top}$, now on top of the stack, by trying all operator identities of Def.~\ref{def:opids}.
If an operator identity applies, $o~o_\textit{top} = \epsilon$, 
then $o_\textit{top}$  
is popped from the stack in the then-branch of the second conditional \texttt{if} (\eg,\ \texttt{hd = 'MX' }$\wedge$\texttt{ top = 'PX'} in the test).
Otherwise, 
the new operator
$o$ is pushed on the stack in the else-branch because the operator pair is not reducible. 
This is the rightmost, innermost reduction strategy 
discussed in Sect.~\ref{sec:pingpong}.
Next, $v$ is pushed and the search continues with $v$ as the new current node, unless $v=\tnode$ and the stack is empty.

If a path exists from $\snode$ to $\tnode$ along which all operators can be reduced to
$\epsilon$, which means the stack is empty (predicate \texttt{bottom} is true) at the target node, the program halts with \texttt{accept}.
This tells us that the protocol is \emph{insecure}. 
The protocol is \emph{secure} if no path from $\snode$ to $\tnode$ is labeled with a reducible word, that is, the input is rejected.
The program defines how to search for an insecure path using a nondeterministic~choice. 
\mypara{Tape Representation} 

The representation of the protocol graphs in Figs.~\ref{fig:FSAprotocolone} and~\ref{fig:FSAprotocoltwothree} on the tape of the pushdown program is shown in Fig.~\ref{fig:FSAtape}.
The node names (\texttt{0}, \texttt{1}, ...) and operator names (\texttt{EX}, \texttt{EY}, ...) are included in the tape symbols, so each takes one position on the tape. The length of the tape is $O(|E|)$ given a graph $G=(V,E)$. The representation of the 18 edges of Protocol 1 takes 56 symbols, the 23 edges of Protocol 2 take 71 symbols, and the 28 edges of Protocol 3 take 86 symbols including the two endmarkers. Names are introduced for intermediate nodes because each edge can only be labeled by a single operator. For example, the initial word of user $X$ in Protocol 3 between source node \texttt{0} and target node \texttt{1} takes three labeled edges with two intermediate nodes, which we call \texttt{8} and~\texttt{9}.

(If node names are encoded as numbers 
in a fixed set of tape symbols (\eg, $\texttt{0}, \texttt{1}$), the tape length would be $O(|E|\cdot\log\:|V|)$, and the program would need to be adapted to deal with the encoding and separators between the edges on the tape, \eg, edge $(3, E_X, 5) = \texttt{1 1 EX 1 0 1 \$}$.)

\begin{figure}[t]
\begin{framedstuff}
\centering
\begin{minipage}[t]{151mm}
\begin{codesize}
\small %
\begin{alltt}
\textrm{Protocol 1:}
> 0 EY 1                                                \textrm{initial word of trusted user X (\mymath{\alpha\sb{1}})}
  1 EX 1 1 PX 1 1 MX 1 1 EY 1 1 PY 1 1 MY 1             \textrm{words of attacker Z (\mymath{\Sigma\sb{Z}\sp{*}})}
  1 EZ 1 1 PZ 1 1 MZ 1 1 M  1 1 DZ 1
  1 DX 2 2 EY 1 2 EZ 1 1 DY 3 3 EX 1 3 EZ 1 <           \textrm{words of trusted users X,Y (\mymath{\alpha\sb{2}})}

\textrm{Protocol 2:}
> 0 PX 8 8 EY 1                                         \textrm{initial word of trusted user X (\mymath{\alpha\sb{1}})}
  1 EX 1 1 PX 1 1 MX 1 1 EY 1 1 PY 1 1 MY 1             \textrm{words of attacker Z (\mymath{\Sigma\sb{Z}\sp{*}})}
  1 EZ 1 1 PZ 1 1 MZ 1 1 M  1 1 DZ 1
  1 DX 2 2 MY 5 5 EY 1 2 MZ 6 6 EZ 1                    \textrm{words of trusted users X,Y (\mymath{\alpha\sb{2}})}
  1 DY 3 3 MX 4 4 EX 1 3 MZ 7 7 EZ 1 <

\textrm{Protocol 3:}
> 0 EY 8 8 PX 9 9 EY 1                                  \textrm{initial word of trusted user X (\mymath{\alpha\sb{1}})}
  1 EX 1 1 PX 1 1 MX 1 1 EY 1 1 PY 1 1 MY 1             \textrm{words of attacker Z (\mymath{\Sigma\sb{Z}\sp{*}})}
  1 EZ 1 1 PZ 1 1 MZ 1 1 M  1 1 DZ 1
  1 DX 2 2 MY 4 4 DX 11 11 EY 1 2 MZ 5 5 DX 12 12 EZ 1  \textrm{words of trusted users X,Y (\mymath{\alpha\sb{2}})}
  1 DY 3 3 MX 6 6 DY 10 10 EX 1 3 MZ 7 7 DY 13 13 EZ 1 <
\end{alltt}
\end{codesize}
\end{minipage}
\end{framedstuff}
\mycaption{Tape representation of the protocol graphs in Figs.~\ref{fig:FSAprotocolone} and~\ref{fig:FSAprotocoltwothree} (nodes and operators are represented by tape symbols \texttt{0}, \texttt{1}, ..., \texttt{EX}, \texttt{EY}, ...).}
\label{fig:FSAtape}
\end{figure}%

\mypara{Experiments and Implementation} 

The pushdown program in Fig.~\ref{fig:PDAverify} defines a 1-head, 2-way and nondeterministic pushdown automaton (1-head 2NPDA). This is easy to see because the program has a single head variable (\texttt{hd}), moves in both directions (\texttt{left}, \texttt{right}), and has a nondeterministic choice (\texttt{choice}). 
Because a $k$-head 2NPDA can be simulated in at most $O(n^{3k})$ steps where $n$ is the length of the tape, checking the security of ping-pong protocols by a $1$-head 2NPDA takes at most $O(n^{3})$~steps.

The pushdown simulator~\cite{Glueck:16:pdasim} takes a set of transition rules as the definition of a pushdown automaton. It takes 690~transition rules to define the program shown in Fig.~\ref{fig:PDAverify} as a 1-head 2NPDA with 8 control states, 30 tape symbols, and 30 stack symbols. The number of transition steps and surface configurations (state $\times$ stack-top symbol $\times$ tape symbol) the simulator takes to verify the three Dolev-Yao protocols are listed in Table~\ref{tab:simruntape}.
The simulator looks for a universal solution, exploring all computation sequences leading to an accept (insecure protocol) and not halting at the first accept being found.
Thus, the outcome and the order of the edges on the tape have no significant influence on the performance of the simulation.

\begin{table}[ht]
\caption{Simulation of the pushdown program as 2NPDA with the Dolev-Yao example protocols.}
\begin{center}
\begin{tabular}{@{}lrrrrl@{}}
\hline
\emph{verification} & \emph{edges} & \emph{tape} & 
\emph{configs}      & \emph{steps} & \emph{answer} \\
\hline
Protocol~1 & 18~~ & 56~\, &  584~~ &  8100 & accept (insecure)\\
Protocol~2 & 23~~ & 71~\, &  740~~ &  6031 & reject (secure)\\
Protocol~3 & 28~~ & 86~\, & 1184~~ & 11412 & accept (insecure)\\
\hline
\end{tabular}
\end{center}
\label{tab:simruntape}
\end{table}%

\mypara{Discussion}

The nondeterministic pushdown verifier
is surprisingly simple (in the author's opinion), especially when considering that Dolev and Yao's first algorithm took $O(n^8)$ steps~\cite{DolevYao:81,DolevYao:83}, while the verifier in Fig.~\ref{fig:PDAverify} is guaranteed to take at most $O(n^3)$ steps thanks to fundamental results of 
automata theory.  
The verifier checks whether the intersection of a regular language represented on the tape as FSA (the protocol to verify) and a fixed Dyck-like language containing all canceling (insecure) words is empty. Clearly, the verifier is not limited to protocol graphs constructed by the algorithm~\cite{DolevEvenKarp:82}. 
Any operator-labeled FSA can be placed on the tape and the operator identities tested in the program can be adapted easily to other identities. Thus, any security question that can be captured by an operator-labeled FSA (any regular language) intersected with a Dyck-like language induced by a fixed set of operator identities can be decided by the verifier after adaption to the specific operator identities.

The experiment also supports the proposition that 
programming languages %
can make abstract theoretical results more accessible and applicable~\cite{Jones:97:complexity,Reus:16} 
(other examples are reversible programming languages~\cite{AxGl11FoSSaCS}). The simulation approach is also amenable to optimization by program specialization, which typically reduces the interpretive overhead by an order of magnitude. 
However, a memoizing interpreter for a nondeterministic language may pose additional challenges to non-trivial specialization.
A downside, from a programming perspective, is that the pushdown language has no convenient data structures, just a linear tape with symbols. A search for matching edges needs to cycle over all edges on the tape and to return the head from the right end to the left end.

\section{Protocol Verification by Transformation and Pushdown Simulation}
\label{sec:vptsim}

Verification by general-purpose program transformers is another approach that has been proposed and used successfully to check the security of cryptographic protocols by supercompilation~\cite{AhmedLisitsaNemytykh:13,Nepeivoda:16}.

Typically, the security problem of a protocol (\eg, a ping-pong protocol)
is encoded as a functional program with an additional trace parameter that constrains all nondeterministic choices such that they become deterministic. 
Given a trace, 
the functional program 
maps an initial state into a final state which can then be tested for validity (\eg, whether the state is insecure).
By specializing the program {\wrt} a static (known) initial state and a dynamic (unknown) trace, a supercompiler will explore the control flow of all possible traces for the given initial state.\footnote{The trace program may have additional parameters depending on the particular class of protocols and the desired answer.}
If the program is specialized into a residual program from which 
it is immediately seen
that no trace can steer the residual program into an invalid final state
then the original protocol is considered secure. 

A difficulty with this approach is the preparation of a functional program that specializes well, which may require considerable knowledge about 
supercompilation.
There are obstacles, namely finding the right encoding of the security problem in a universal source language and taming the power of the supercompiler with folding, generalization and other sophisticated optimizations, especially if success depends on the way the program is written.
The nondeterminism inherent in the security problem 
is first mapped 
into a deterministic program
and then reintroduced 
by specializing the deterministic program {\wrt} a dynamic trace, which can make the transformation hard to predict. 
A nondeterministic choice in the original problem does not necessarily correspond to a nondeterministic (dynamic) choice taken by the supercompiler  when exploring all possible control flows in the program. The control over these choices 
is therefore indirect.
On the other hand, case studies~\cite{AhmedLisitsaNemytykh:13,LisitsaNemytykh:07} have shown that 
a supercompiler~\cite{Turchin:86} can solve a variety of 
security problems, including cryptographic ping-pong protocols~\cite{Nepeivoda:16}.

In our case study, the protocol verifier is a pushdown program that explores all paths in a protocol graph by nondeterministic means. The security problem is not encoded as a functional program, but given as input data (Fig.~\ref{fig:FSAtape}) to the verifier (Fig.~\ref{fig:PDAverify}). The verifier is the same for all protocols.
A branching in the graph on the tape is directly modeled by a nondeterministic choice in the pushdown program. The verifier has exactly one 
nondeterministic choice point; all other choices in the program are deterministic (\texttt{if}). The nondeterministic \texttt{choice} is an integral part of the  language semantics  
and gives direct control of the problem-specific nondeterminism inherent in the security problem. The verifier is written in a 
language for which decidability 
is guaranteed,
which is a major advantage of this approach.
To the best of the author's knowledge, this is the first study in which the pushdown programming model has been applied to ping-pong protocols, but the model's practicality for verifying a larger class of protocols has not yet been demonstrated.
At least in principle, the multihead pushdown programs 
can decide any
computationally ``tractable'' verification problem.

Both approaches explore all possible paths of a security problem  %
regardless of its representation.
In one approach, this is achieved by a supercompiler specializing a program representing the problem and building an internal process graph, whereas in the other approach it is achieved by a simulator of pushdown programs using memoization.
The nondeterministic choice is an integral part of the pushdown language semantics, while in the case of 
supercompilation the nondeterminism is induced into the source program by the supercompiler's non-standard ``transformation semantics''~\cite{AbrGlu:00:nsint}, 
not the standard semantics of the  source language. 
Neither approach shows why a protocol is insecure or how to fix it, but a step in this direction has been taken by building attack models~\cite{Nepeivoda:16}.

\begin{table}[t]
\caption{Pushdown simulation and supercompilation characteristics.}
\begin{center}
\begin{tabular}{|c|c|c|}
\hline
\emph{method}          & SIM                    & SCP \\
\hline
\emph{approach}        & interpretation         & transformation    \\
\hline
\emph{sound    answer} & yes                    & yes             \\
\emph{complete answer} & yes                    & case by case    \\
\hline
\emph{time  complexity class} & $=$ polynomial         & exponential?    \\
\emph{space complexity class} & $\subseteq$ polynomial & unknown         \\
\hline
                       & subuniversal           & universal       \\
\emph{source language} & nondeterministic       & deterministic   \\
                       & tail-recursive         & recursive       \\
\hline
\end{tabular}
\end{center}
\label{tab:SIMSCPprop}
\end{table}%

Table~\ref{tab:SIMSCPprop} summarizes the two approaches. The entries in the table are for a general-purpose supercompiler~\cite{Turchin:86}, bearing in mind there are different supercompiler variants (\eg, \cite{LisitsaNemytykh:07,Nepeivoda:16,sorm99:intro}).
The simulator (SIM) interprets a pushdown program for a given 
protocol graph (sequence of edges) on the 
input tape and the supercompiler (SCP) transforms the program representation of the security problem into a residual program. 
Provided that SIM is correct, the answer (accept/reject) is sound and complete (multihead 2NPDA are decidable). Provided that SCP is correct, 
the generated residual program is a sound answer,
but depending on the particular protocol encoding, the residual program may not always answer the security question (\eg, due to overgeneralization it contains valid and invalid final states even though the protocol is secure, or due to infinite specialization no program is generated at all). Thus, completeness of the answer is marked as `case by case' in the table. (An exception is 
the verification of ping-pong protocols
which was shown to be decidable for all multi-party ping-pong protocols~\cite{Nepeivoda:16}.)
Exponential time complexity was conjectured for ping-pong protocol verification by supercompilation~\cite{Nepeivoda:16}, while the space complexity class is unknown. The time and space complexity classes of multihead 2NPDA are guaranteed~\cite{WagnerWechsung:86}. 
The source language of SIM is a subuniversal (not Turing-complete), nondeterministic and tail-recursive 
pushdown language (Sect.~\ref{sec:ndetpgmoverview}), while the source language of SCP is a universal (Turing-complete), deterministic and recursive first-order functional language. The deterministic and nondeterministic property of their source languages is also indicated in~Fig.~\ref{fig:VERIFYapproaches}.

\section{Related Work}
\label{sec:related}

It has been know for several decades that nondeterministic programs are well suited for combinatorial search problems, and in many cases even easier to write than deterministic ones~\cite{Floyd:67}.
An obstacle is the effective control of the often exponential-time complexity of straightforward runs and
nonterminating computation paths.
For two-way nondeterministic pushdown automata, a polynomial-time and terminating bottom-up simulation algorithm could be given~\cite{AHU:68}. Another simulation algorithm~\cite{Glueck:16:pdasim} follows top-down, all reachable computation paths, as does the one for two-way deterministic pushdown automata~\cite{Jones:77}. 
In these algorithms, exponential time is converted into polynomial time by 
sharing computations.
A linear-time simulation by instrumented two-way deterministic pushdown programs is given in~\cite{Mogensen:94}.
Certain methods of 
model checking~\cite{ABE:16:handbook} also make use of pushdown systems. A classic case where investigations in pure theory of pushdown automata led to a practically significant algorithm is pattern matching~\cite{KnuMorPra:77}. 
The case study presented here appears to be one of the first to show how a nondeterministic pushdown language can be used as a 
decidable 
programming model for protocol verification.

A series of case studies examined the verification of protocols by program transformation (\eg,~\cite{LisitsaNemytykh:07}), and in particular 
cryptographic protocols by supercompilation (\eg,~\cite{AhmedLisitsaNemytykh:13,Nepeivoda:16}). A few orthogonal supercompilation principles~\cite{Turchin:86} were shown to solve a number of seemingly different verification problems.
Related
approaches for specific verification problems have been investigated (\eg, \cite{GlueckLeuschel:99:PSI,Klimov:12}) and 
comparable results can be conjectured (\eg,~\cite{FKG:01:NGC}).
Interpreters have been used to improve the transformation of programs~(\eg,~\cite{Glueck:94:JFP,GlueckJoergensen:94:SAS}), which is another way to factorize supercompilation-based verification.
The literature on transformation-based verification is larger than the one cited here, including approaches based 
on unfold/fold rules (\eg,~\cite{FioravantiPettorossiProietti:02}).
The present study examined a related, but different programming language solution, namely nondeterministic programs in a decidable computation model with guaranteed resource bounds. Related principles underlying both  
approaches, verification by transformation and interpretation, were discussed above (Sect.~\ref{sec:vptsim}). 
Extensions of logic languages with tabulation can ensure 
termination and 
optimal known complexity for queries to a large class of practical 
programs~\cite{SwiftWarren:12}.

\section{Conclusions and Further Work}
\label{sec:conclusion}

This study broadens previous studies on the verification of security by program transformation in that another programming language approach, namely program interpretation, is used. We confirmed that the security of a well-studied class of cryptographic protocols
can be verified by a 1-head 2NPDA. The interpretive approach used in this experiment considerably simplified 
the verification,
by separating nondeterministic pushdown logic 
from control 
concerns,
which shows again the power of a declarative style of programming.

Program transformation and interpretation are two sides of the same coin, and we
identified principles that both verification approaches share 
(\eg, explore all possible paths of a security problem).
Also noteworthy is how abstract results from automata theory can be applied to practical problems when combined with a programming language approach, and that this can yield more natural and simple solutions.
This study is one of the few examples in the literature where pushdown automata have been used to answer questions other than those of formal language theory. 
This situation is perhaps surprising because the multihead 2NPDA programming model is 
equivalent to the class of polynomial-time algorithms 
and decidable within guaranteed time and space bounds determined by the number of heads.

Though we showed how a
class of cryptographic protocols can be verified by nondeterministic programming, further work is needed before a more complete picture emerges as to the practicality of the interpretive pushdown approach. The pushdown computation model is subuniversal, and thus cannot be expected to be capable of solving all verification problems in reach of a general-purpose program transformer, such as a supercompiler using sophisticated generalization techniques and capable of generating complex recursive programs as answers. On the other hand, the multihead nondeterministic pushdown model is theoretically powerful enough to decide 
all
polynomial-time verification problems, but whether this is as 
straightforward as in the case
of ping-pong protocols, only further investigations will show. Verification of multi-party extended protocols~\cite{Nepeivoda:16} in further studies is warranted.

From a programming perspective, the 
``machine-code''  
transition rules of classic presentations of pushdown automata are too low-level.
A step towards a more user-friendly abstraction was undertaken in this paper by employing an imperative flowchart language with deterministic and nondeterministic control-flow operators. Still, the language inherits the linear  input tape from automata theory. More practical data structures and languages abstractions could be considered, {\eg}\ arrays and index calculations~\cite{Mogensen:94}, tree-structured data 
or cons-free functional programming languages~\cite{Jones:97:complexity}.

\mypara{Acknowledgements}

Thanks to Antonina Nepeivoda for a concise explanation of the Dolev-Yao model and to the anonymous reviewers for their constructive feedback.

\end{document}

